\documentclass[usenatbib]{mn2e}
\usepackage{graphicx}
\usepackage{amssymb}
\usepackage{wasysym}
\usepackage[dvips]{hyperref}

\def\rpd{\hbox{rad\,d$^{-1}$}}
\def\chisq{\hbox{$\chi^2$}}
\def\chisqr{\hbox{$\chi^2_r$}}
\def\msun{\hbox{${\rm M}_{\odot}$}}

\def\rsun{\hbox{${\rm R}_{\odot}$}}

\def\rstar{\hbox{$R_{\star}$}}

\def\tq{\hbox{$T_{\rm q}$}}
\def\ts{\hbox{$T_{\rm s}$}}

\def\sn{\hbox{S/N}}

\def\kms{\hbox{km\,s$^{-1}$}}
\def\vsini{\hbox{$v\sin i$}}

\def\ptt{\hbox{$10^{-4} I_{\rm c}$}}

\def\degr{\hbox{$^\circ$}}

\def\Prot{\hbox{$P_{\rm rot}$}}
\def\Omeq{\hbox{$\Omega_{\rm eq}$}}
\def\dOm{\hbox{$d\Omega$}}

\newcommand{\aap}{A\&A}

\newcommand{\apj}{ApJ}

\newcommand{\apss}{Ap\&SS}

\newcommand{\mnras}{MNRAS}

\newcommand{\caii}{\hbox{Ca$\;${\sc ii}}}

\begin{document}

\title[The stable magnetic field of V374~Peg] 
{The stable magnetic field of the fully-convective star 
V374~Peg\thanks{Based on observations obtained at the Canada-France-Hawaii Telescope (CFHT) 
which is operated by the National Research 
Council of Canada, the Institut National des Science de l'Univers of the Centre 
National de la Recherche Scientifique of France, and the University of Hawaii.} }

\makeatletter

\def\newauthor{%
  \end{author@tabular}\par
  \begin{author@tabular}[t]{@{}l@{}}}
\makeatother
 
\author[J. Morin et al.] {\vspace{1.7mm}
J.~Morin$^1$\thanks{E-mail: 
jmorin@ast.obs-mip.fr (JM);
donati@ast.obs-mip.fr (J-FD);
thierry.forveille@obs.ujf-grenoble.fr (TF);
xavier.delfosse@obs.ujf-grenoble.fr (XD);
wolfgang.dobler@ucalgary.ca (WD); 
petit@ast.obs-mip.fr (PP);
mmj@st-andrews.ac.uk (MMJ);
acc4@st-andrews.ac.uk (ACC);
albert@cfht.hawaii.edu (LA); 
manset@cfht.hawaii.edu (NM); 
dintrans@ast.obs-mip.fr (BD);
chabrier@ens-lyon.fr (GC);
valenti@stsci.edu (JV)},
J.-F.~Donati$^1$, T.~Forveille$^2$, X.~Delfosse$^2$, W.~Dobler$^3$, P.~Petit$^1$, \\ 
\vspace{1.7mm}
{\hspace{-1.5mm}\LARGE\rm
M.M.~Jardine$^4$, A.C.~Cameron$^4$, L.~Albert$^5$, N.~Manset$^5$, B.~Dintrans$^1$, } \\ 
\vspace{1.7mm}
{\hspace{-1.5mm}\LARGE\rm
G.~Chabrier$^6$, J.A.~Valenti$^7$ } \\
$^1$ LATT--UMR~5572, CNRS et Univ.\ P.~Sabatier, 14 Av.\ E.~Belin, F--31400 Toulouse, France\\
$^2$ LAOG--UMR~5571, CNRS et Univ.\ J.~Fourier, 31 rue de la Piscine, F--38041 Grenoble, France \\
$^3$ Dept.~of Physics and Astronomy, Univ.\ of Calgary, Calgary, Alberta T2N~1N4, Canada \\
$^4$ School of Physics and Astronomy, Univ.\ of St~Andrews, St~Andrews, Scotland KY16 9SS, UK \\
$^5$ CFHT, 65-1238 Mamalahoa Hwy, Kamuela HI, 96743 USA \\ 
$^6$ CRAL, CNRS--UMR~5574, Ecole Normale Sup\'erieure de Lyon, 46 all\'ee d'Italie, F--69362 Lyon, France \\ 
$^7$ STScI, 3700 San Martin Drive, Baltimore MD, 21218 USA } 

\date{\today,~ $Revision: 1.7 $}
\maketitle
 
\begin{abstract} 
We report in this paper phase-resolved spectropolarimetric observations of the rapidly-rotating 
fully-convective M4 dwarf V374~Peg, on which a strong, mainly axisymmetric, large-scale poloidal 
magnetic field was recently detected.  In addition to the original data set secured in 2005 August, 
we present here new data collected in 2005 September and 2006 August.  

From the rotational modulation of unpolarised line profiles, we conclude that starspots are 
present at the surface of the star, but their contrast and fractional coverage are much lower than 
those of non-fully convective active stars with similar rotation rate.  
Applying tomographic imaging on each set of circularly polarised profiles separately, we find that 
the large-scale magnetic topology is remarkably stable on a timescale of 1~yr;  repeating the 
analysis on the complete data set suggests that the magnetic configuration is sheared by 
very weak differential rotation (about 1/10th of the solar surface shear) and only 
slightly distorted by intrinsic variability.  

This result is at odds with various theoretical predictions, suggesting that dynamo fields of 
fully-convective stars should be mostly non-axisymmetric unless they succeed at triggering significant 
differential rotation.  
\end{abstract}


\section{Introduction} 

Most cool stars exhibit demonstrations of activity such as cool surface spots, coronal activity or 
flares, usually attributed to magnetic fields.  The current understanding is that these 
magnetic fields are generated at the surfaces and in the convective envelopes of cool stars 
through dynamo processes, involving cyclonic motions of plasma and rotational 
shearing of internal layers. In partly-convective Sun-like stars, dynamo processes are thought 
to concentrate where rotation gradients are supposedly largest, i.e., at the interface layer between 
their radiative cores and convective envelopes.  Cool stars with masses lower than 0.35~\msun\ being 
fully convective \citep{Chabrier97}, they lack the thin interface layer presumably hosting 
solar-type dynamo processes and are thought to rotate mainly as rigid bodies \citep{Barnes05, 
Kuker05};  yet they are both very active \citep[e.g.,][]{Delfosse98} and 
strongly magnetic \citep{Saar85, Johns96, Reiners07}.  

This led theoreticians to propose that the intense activity and strong magnetism of fully convective 
dwarfs may be due to non-solar dynamo processes, in which cyclonic convection and turbulence play the 
main roles while differential rotation contributes very little \citep[e.g.,][]{Durney93};  the exact 
mechanism capable of producing such fields is however still a debated issue.
The most recent numerical dynamo simulations suggest that these stars 
are actually capable of producing large-scale magnetic fields, although they disagree on the actual 
properties of the large scale field.  While some find that fully-convective stars rotate as 
solid bodies and host purely non-axisymmetric large-scale fields \citep{Kuker05, Chabrier06}, 
others diagnose that these stars should succeed at triggerring differential rotation and thus 
produce a net axisymmetric poloidal field \citep[e.g.,][]{Dobler06}.

Existing observational data on fully convective dwarfs do not completely agree with any of these 
models.  Observations of low-mass stars indicate that surface differential rotation vanishes with
increasing convective depth, so that fully-convective stars rotate mostly as solid bodies 
\citep{Barnes05}.  At first sight, this result confirms nicely the predictions of \citet{Kuker05}, 
leading one to expect fully-convective stars to host purely non-axisymmetric large-scale fields 
\citep{Chabrier06}.  
However, spectropolarimetric data of the rapidly rotating M4 dwarf V374~Peg (G~188-38, GJ~4247, HIP~108706) collected with ESPaDOnS 
for almost three complete rotation periods over nine rotation cycles \citep[hereafter D06]{Donati06} 
demonstrated that this fully-convective star hosts a strong mostly-axisymmetric poloidal field despite 
rotating almost rigidly, in contradiction with theoretical expectations.  

In this new paper, we present and analyse spectropolarimetric data of V374~Peg collected with ESPaDOnS 
at three different epochs, 2005 August (i.e., those used in D06), 2005 September (providing only partial 
coverage of the rotation cycle) and 2006 August (providing full, though not very dense, coverage of the 
rotation cycle).  
First, we briefly describe the new observations and the Zeeman detections we obtained.  We then apply 
tomographic imaging to our data sets and characterise the spot distributions and magnetic topologies at 
the surface of V374~Peg in both 2005 and 2006;  from the observed temporal evolution of the magnetic 
topology, we also estimate both the differential rotation and the long-term magnetic stability of V374~Peg.  
Finally, we discuss the implications of our results for our understanding of the dynamo processes 
operating in fully-convective stars. 

\begin{table*}
\begin{center}
\caption[]{Journal of observations.  Subsets of observations collected sequentially on the same night (with 
no time gaps) are listed on a single line.  Columns 1--7 list the UT date, the number of observations in 
each subset, the heliocentric Julian date, the UT time, the exposure time, the peak signal to noise ratio 
(per 2.6~\kms\ velocity bin) and the rms noise level (relative to the unpolarised continuum level and per 
1.8~\kms\ velocity bin) in the average circular polarization profile produced by Least-Squares Deconvolution 
(see text). The rotational cycle $E$ from the ephemeris of equation (\ref{eq:eph}) is given in column 8. Whenever several observations are present in a subset, cols.~3, 4 and 8 list the values 
corresponding to the first and last subset observations, whereas cols.~6 and 7 mention the minimum and 
maximum values reached within the subset.  }
\begin{tabular}{@{}cccccccc}
\hline
Date & $n_{\rm obs}$ & HJD          & UT      & $t_{\rm exp}$ & \sn\  & $\sigma_{\rm LSD}$ & Cycle          \\
           & & (2,453,000+) & (h:m:s) &   (s)         &       &   (\ptt) \\
\hline
   2005   & & & & & & & \\
\hline
Aug 19 & 6  & 601.78613--601.87607 & 06:43:29--08:52:59 & 4 $\times$ 300.0 & 145--170 & 6.0--7.7 & 0.00--0.20 \\
Aug 19 & 12 & 601.94413--602.13790 & 10:30:60--15:10:01 & 4 $\times$ 300.0 & 99--172 & 5.6--11.3 & 0.35--0.79 \\
Aug 21 & 23 & 603.75042--604.19199 & 05:52:00--15:15:52 & 4 $\times$ 300.0 & 137--208 & 5.7--7.8 & 4.41--5.40 \\
Aug 23 & 23 & 605.73892--606.13069 & 05:35:25--14:59:33 & 4 $\times$ 300.0 & 125--176 & 6.0--8.8 & 8.87--9.75 \\
\hline
Sep 08 & 6  & 621.94402--622.02801 & 10:30:49--12:31:46 & 4 $\times$ 300.0 & 135--161 & 7.1--9.2 & 45.24--45.43 \\
\hline
   2006   & & & & & & \\
\hline
Aug 05 & 2 & 952.92084--952.93771 & 09:58:04--10:22:22 & 4 $\times$ 275.0 & 172--181 & 5.7--5.8 & 788.00--788.04 \\
Aug 05 & 1 & 953.05250 & 13:07:39 & 4 $\times$ 275.0 & 180 & 5.6 & 788.30 \\
Aug 07 & 2 & 955.02780--955.04651 & 12:31:58--12:58:55 & 4 $\times$ 300.0 & 192--192 & 5.3--5.3 & 792.73--792.78 \\
Aug 08 & 2 & 955.86045--955.87780 & 08:30:57--08:55:55 & 4 $\times$ 300.0 & 163--182 & 5.5--6.4 & 794.60--794.64 \\
Aug 08 & 2 & 956.02162--956.03903 & 12:23:01--12:48:05 & 4 $\times$ 300.0 & 194--198 & 5.2--5.4 & 794.96--795.00 \\
Aug 09 & 2 & 956.82353--956.84159 & 07:37:44--08:03:44 & 4 $\times$ 300.0 & 185--191 & 5.2--5.5 & 796.76--796.80 \\
Aug 09 & 2 & 957.02738--957.04305 & 12:31:15--12:53:50 & 4 $\times$ 260.0 & 182--187 & 5.3--5.4 & 797.22--797.26 \\
Aug 10 & 2 & 957.84510--957.86535 & 08:08:44--08:37:54 & 4 $\times$ 360.0 & 212--222 & 4.6--4.6 & 799.06--799.10 \\
Aug 10 & 2 & 958.03059--958.05071 & 12:35:50--13:04:49 & 4 $\times$ 360.0 & 191--208 & 4.7--5.1 & 799.47--799.52 \\
Aug 11 & 2 & 959.03187--959.05226 & 12:37:38--13:06:59 & 4 $\times$ 360.0 & 215--220 & 4.7--4.7 & 801.72--801.76 \\
Aug 12 & 2 & 960.03589--960.05609 & 12:43:23--13:12:27 & 4 $\times$ 360.0 & 205--205 & 4.8--4.9 & 803.97--804.02 \\
\hline
\end{tabular}
\end{center}
\label{tab:log}
\end{table*}

\section{Observations}
\label{sec:obs}

Spectropolarimetric observations of V374~Peg were collected in 2005 August (hereafter Aug05), 2005 September 
(hereafter Sep05) and 2006 August (hereafter Aug06) with the ESPaDOnS spectropolarimeter 
\citep[Donati et al, in preparation]{Donati97, Donati06c}  
at the Canada-France-Hawaii Telescope (CFHT).  The ESPaDOnS spectra span the entire optical domain 
(from 370 to 1000~nm) at a resolving power of about 65,000.  During the Aug05 and Aug06 runs, V374~Peg was 
observed several times per night to ensure correct sampling of the rotational cycle (0.4456~d, D06).  A 
total of 91 circular-polarization (Stokes $V$) observations were collected, each consisting of four 
individual subexposures taken in different polarimeter configurations (to suppress all spurious polarisation 
signatures at first order, \citealt{Donati97}). 

Data reduction was performed using Libre~ESpRIT \citep[Donati et al 2007, in preparation]{Donati97}, a 
fully automatic reduction package installed at CFHT and achieving optimal extraction of ESPaDOnS spectra.  
The peak signal-to-noise ratio (\sn) per 2.6~\kms\ velocity bin in the recorded spectra ranges from 99 
to 222, depending mostly on weather conditions and exposure time (see Table~\ref{tab:log}).  
The Aug05 and Aug06 data sets respectively include 64 and 21 
spectra;  4 spectra in the Aug05 data set are significantly affected by flaring and were discarded for 
the present study.  Along with the 6 spectra collected in Sep05, the complete data set 
used for this study includes 87 spectra.  

In order to increase further the quality of our data set, we used
Least-Squares Deconvolution \citep[LSD,][]{Donati97} to extract the
polarimetric information from most photospheric spectral lines.  The line
list required for LSD was computed with an Atlas9 model atmosphere
\citep{Kurucz93} matching the properties of V374~Peg and includes about
5~000 spectral features.  We finally obtain LSD Stokes $V$ profiles with
relative noise levels of about 0.05\% (in units of the unpolarised
continuum), corresponding to a typical multiplex gain of about
10\footnote{This multiplex gain refers to the peak \sn\ achieved in the
spectrum, i.e., around 850~nm in the case of M dwarfs and ESPaDOnS
spectra. Previous studies, reporting higher gains
\citep[e.~g.][]{Donati97}, were based on spectra at shorter
wavelengths (ranging from about 500 to 700~nm), where the line density at
peak \sn\ is comparatively higher than in the present case.} with respect
to individual spectra. A typical example is shown in Fig.~\ref{fig:lsd}.  

Zeeman signatures are detected in nearly all the spectra, with amplitudes of up to 
0.5\% (see Fig.~\ref{fig:lsd}) and exhibit clear temporal variations.   
%% (see Fig.~\ref{fig:spec_V}).   
The redward migration of the detected Zeeman signatures is particularly obvious in the Aug05 data set 
(see Sec.~\ref{sec:resV}),  demonstrating unambiguously that the observed
variability is due to 
rotational modulation.  
Stokes $I$ profiles only show a low (though clear) level of variability, with photospheric cool spots 
causing moderate line distorsions (see Sec.~\ref{sec:modI}).  
%% (see Fig.~\ref{fig:spec_I}). 

In the following, all data are phased according to the following ephemeris (shifted by 0.01 phase with respect to that of D06):
\begin{equation}
\mbox{HJD} = 2453601.78613 + 0.4456 E.
\label{eq:eph}
\end{equation}

\begin{figure}
\center{\includegraphics[scale=0.41,angle=-90]{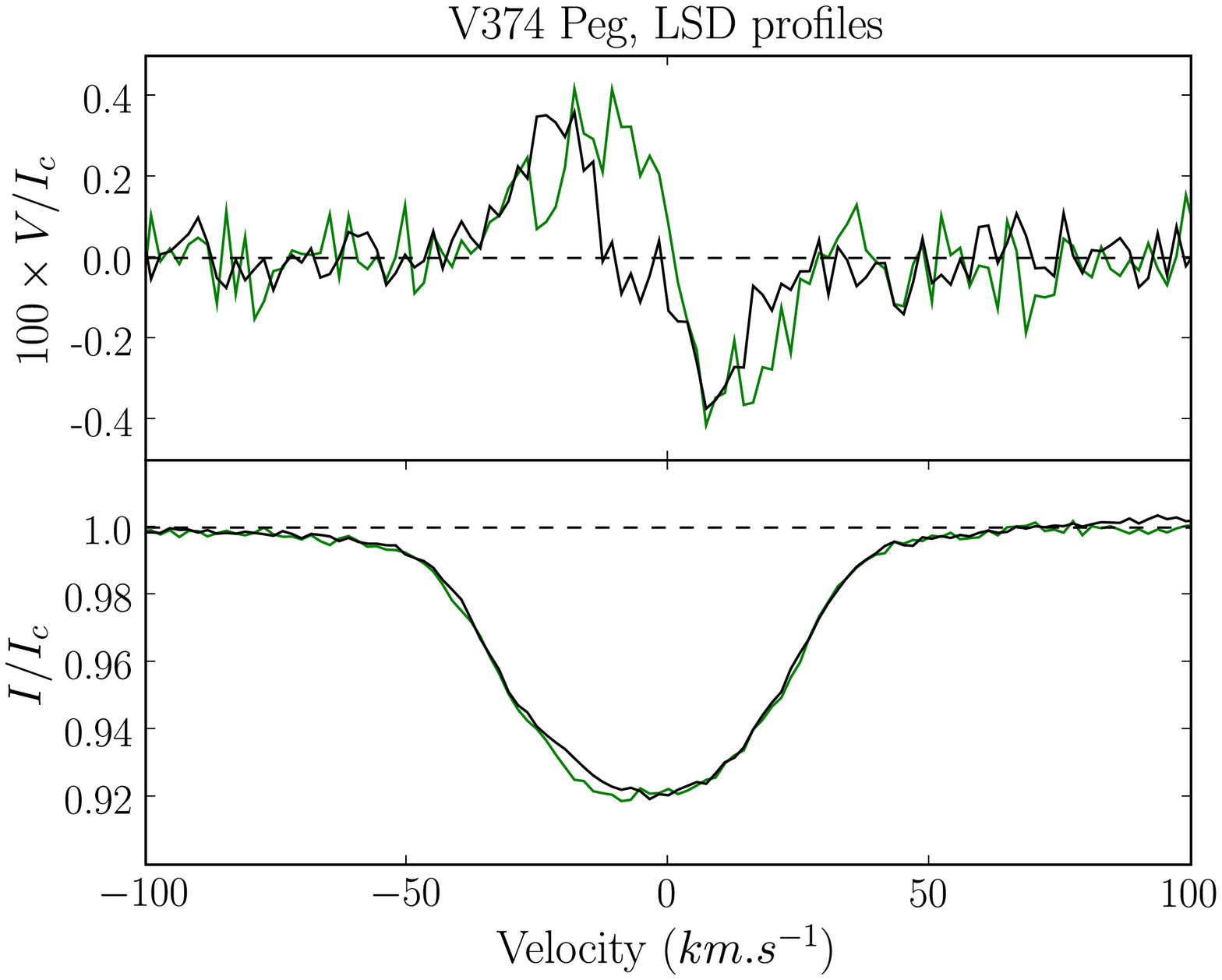}} 
\caption[]{Typical examples of LSD Stokes $V$ and $I$ profiles of V374~Peg
(top and bottom curves respectively) on 2006 August~08 (green line) and
2006 August~10 (black line).}
\label{fig:lsd}
\end{figure}

%% \begin{figure*}
%% \center{\includegraphics[scale=0.9]{fig/v374peg_I_0506_spec.ps}} 
%% \caption[]{LSD Stokes $I$ spectra of V374~Peg.  The rotational phase and cycle of each observation are 
%% written next to each profile. Aug05, Sep05 and Aug06 data correspond to rotation cycles of about $-185$, 
%% $-145$ and 605 respectively.  }
%% \label{fig:spec_I}
%% \end{figure*}

%% \begin{figure*}
%% \center{\includegraphics[scale=0.9]{fig/v374peg_V_0506_spec.ps}} 
%% \caption[]{Same as Fig.~\ref{fig:spec_I} for the LSD Stokes $V$ profiles.  
%% A $3\sigma$ error bar is plotted on the left-hand side of each profile.  } 
%% \label{fig:spec_V}
%% \end{figure*}

\section{Modelling intensity profiles}

\subsection{Model description}
\label{sec:modI}
We use Doppler Imaging (DI) to convert time-resolved series of Stokes $I$ LSD profiles into maps of 
cool spots at the surface of V374~Peg.  Thanks to the Doppler effect, cool spots on the photosphere of 
a rapidly rotating star produce profile distorsions in spectral lines whose locations strongly correlate 
with the spatial positions of the parent spots; in this respect, spectral lines of a spotted star can be viewed 
as one-dimensional (1D) image, resolved in the direction perpendicular to both the stellar rotation axis and 
the line of sight.  By looking at how these 1D maps are periodically modulated as the star rotates enables 
one to recover a 2D map of the stellar photosphere \citep[e.g.,][]{Vogt87}.  Since this inversion problem 
is partly ill-posed, one needs to implement additional constraints to stabilise the imaging process, e.g., 
by selecting the image having the lowest information content using the maximum entropy algorithm of 
\citet{Skilling84}.  More details about the principles and performance DI can be found in \citet{Vogt87}.

We implement this imaging process by dividing the surface of the star into a grid of 3,000 elementary cells.  
We describe the local line profile at each grid point of the stellar surface with the 2-component model of 
\citet{Cameron92}, i.e., as a linear combination between two reference profiles, one representing the quiet 
photosphere (at temperature \tq) and one representing cool spots (at temperature \ts), both Doppler shifted 
by the local line-of-sight rotation velocity of the corresponding grid cell.  The image quantity we reconstruct 
is the local spottedness at the surface of the star, i.e., the amount by which both reference profiles are 
mixed together, varying between 0 (no spot) and 1 (spots only).  The algorithm aims at reconstructing the image with 
minimum spot coverage at the surface of the star, given a certain quality of the fit to the data.  

More specifically, we assume that both reference profiles  are equal and
only differ by their relative continuum levels (set to their respective
blackbody fluxes at the mean line wavelength). We further assume
$\tq=3200$~K (i.e., a typical surface temperature for a M4 star) and
$\ts=2800$~K (i.e., a low spot-to-photosphere temperature contrast in
agreement with the findings of \citealt{Berdyugina05}).  For the assumed
reference profile, we can use either the LSD profile of the very slowly
rotating inactive M dwarf Gl~402 \citep{Delfosse98}, 
or a simple gaussian profile with similar full-width-at-half-maximum
(FWHM, set to 9.25~\kms\ including instrumental broadening) and equivalent
width.  Both options yield very similar results, demonstrating that the
exact shape of the assumed local profile has very minor impact on the
reconstructed images, as long as the projected rotation velocity of the 
star is much larger than the local profile.  We further assume that
continuum limb-darkening varies linearly with the cosine of the limb angle
(with a slope of $u = 0.6964$, \citealt{Claret04});  using a quadratic 
(rather than linear) dependence produces no visible change in the result.

The angle $i$ of the stellar rotation axis to the line-of-sight is taken to be $70\degr$ (e.~g., D06).  Given the projected equatorial velocity \vsini\ of V374~Peg (about 36.5~\kms, see below) and the width of 
the local profile (9.25~\kms, see above), the number of spatially resolved elements across the equator 
of the star is about 25, which is equivalent to a longitude resolution of 15\degr\ or 0.04 rotation cycle.  
Using 3,000 grid cells at the surface of the star (112 across the 
equator) is therefore perfectly adequate for our needs.  

\subsection{Results}
\label{sec:resI}

\begin{figure*}
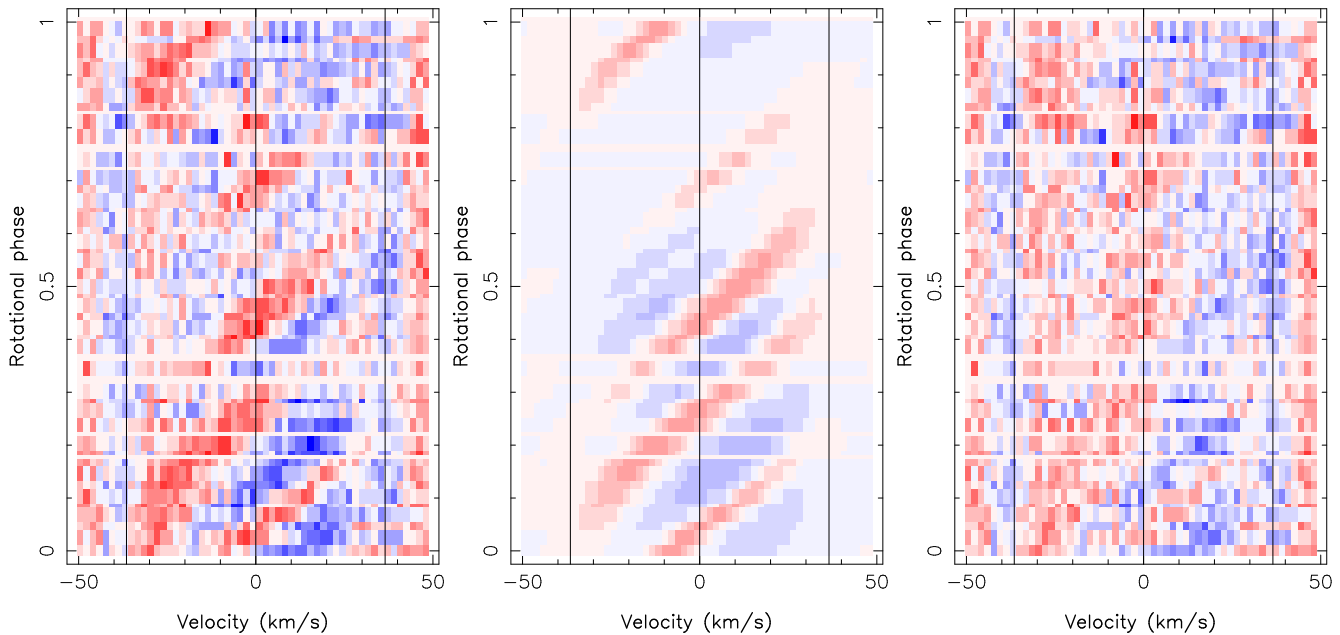

 \center{
 \includegraphics[width=5.8cm]{fig/dyni_aug05b_lpgl402_d.ps}
 \includegraphics[width=5.8cm]{fig/dyni_aug05b_lpgl402_50_r.ps}
 \includegraphics[width=5.8cm]{fig/dyni_aug05b_lpgl402_50_res.ps}}
 \caption[]{Observed (left), modelled (middle) and residual (right) Stokes $I$ dynamic spectra of V374~Peg for the Aug05 data set (60 spectra). A purely rotational profile is subtracted from all spectra to emphasize the image contrast. The fit corresponds to a \sn\ level of $700$. The colour scale ranges from --0.5\% to 0.5\% of the continuum level, red and blue respectively standing for a lack and an excess of absorption with respect to the synthetic purely rotationnal profile.}
 \label{fig:specdynI05}
\end{figure*}

\begin{figure*}
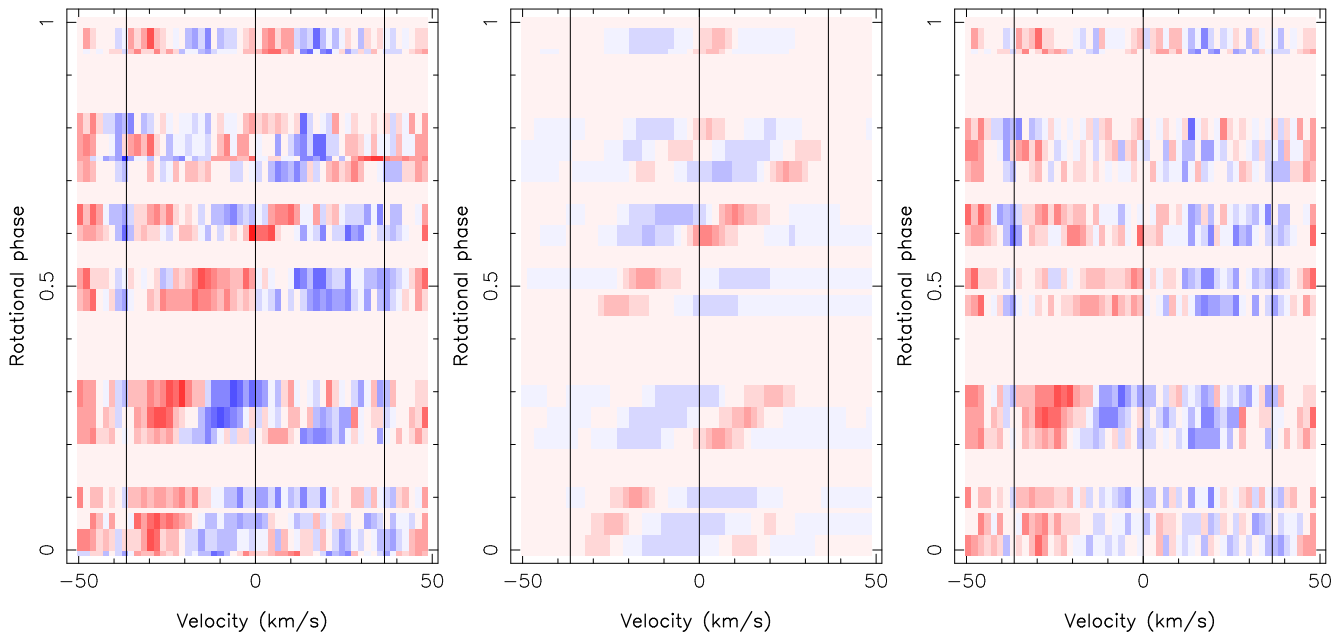

 \center{
 \includegraphics[width=5.8cm]{fig/dyni_aug06_lpgl402_d.ps}
 \includegraphics[width=5.8cm]{fig/dyni_aug06_lpgl402_40_r.ps}
 \includegraphics[width=5.8cm]{fig/dyni_aug06_lpgl402_40_res.ps}}
 \caption[]{Same as Fig.~\ref{fig:specdynI05} for the Aug06 data set (21 spectra).}
 \label{fig:specdynI06}
\end{figure*}

\begin{figure*}
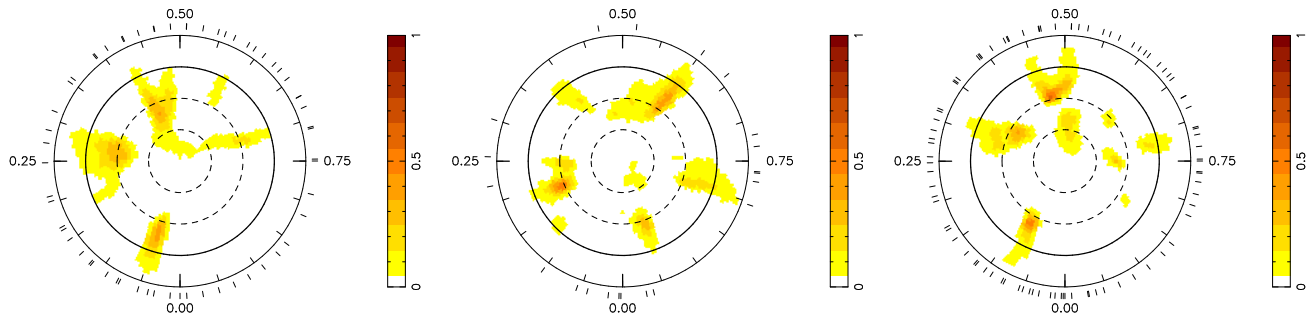

 \center{
 \hbox{\includegraphics[scale=0.33]{fig/v374peg_I_aug05b_lpgl402.ps}\hspace{3mm}
       \includegraphics[scale=0.33]{fig/v374peg_I_aug06_lpgl402.ps}\hspace{3mm}
       \includegraphics[scale=0.33]{fig/v374peg_I_0506_lpgl402.ps}}}
 \caption[]{Doppler maps of the spottedness at the surface of V374~Peg reconstruct, as derived from the 
Aug05 (left panel) and Aug06 (middle panel) data sets. The last image
(right) was recovered from the complete dataset (Aug05, Sep05, and Aug06
altogether), it represents the starspots configuration at an intermediate
epoch (cycle 194) and takes into account the adjustment of the period and
the inclusion of differential rotation (see Sec. 5).  The star is shown in
flattened polar projection down to latitudes of $-30\degr$, with the
equator depicted as a bold circle and parallels as dashed circles.  Radial
ticks around each plot indicate phases of observations.} 
 \label{fig:di_maps}
\end{figure*}

DI is very sensitive to the assumed \vsini\ \citep[e.g.,][]{Vogt87};  small errors are known to generate  
specific belt-like artifacts around the stellar equator in the reconstructed image.  We derive the optimum 
\vsini\ by minimising the reconstructed spot coverage at a given reduced chi-square (hereafter \chisqr) 
fit to the data, and find 
values of $36.5\pm0.4$~\kms\ and $36.7\pm0.4$~\kms\ for the Aug05 and Aug06 data sets respectively, 
when using the gaussian local profile model.  Slightly different values
(all lying with 0.5~\kms\ of the previous estimates) are obtained when
using the LSD profile of Gl~402 as local profile, or when using
different models of the continuum limb darkening.  Taking into account all
potential sources of systematic errors, we 
obtain that the absolute accuracy to which \vsini\ is determined is about
1~\kms;  in the following, 
we assume $\vsini=36.5$~\kms.  

Aug05 and Aug06 datasets can be fitted at \sn\ levels of 715 and 800 respectively. 
In particular, the distorsions travelling from the blue to the red profile wing are well reproduced 
(see Fig.~\ref{fig:specdynI05} and \ref{fig:specdynI06}).  We note small systematic residual discrepancies 
between the observed and modelled profiles, showing up as vertical bands in the far line wings (see 
Fig.~\ref{fig:specdynI05} and \ref{fig:specdynI06}).  These differences cannot be removed by adjusting 
\vsini\ or $u$ and do not affect the imaging of starspots.  Another residual discrepancy is visible in 
the blue profile wing around phase 0.27 in the Aug06 data set (see Fig.~\ref{fig:specdynI06}, right panel).  
This discrepancy corresponds to a feature visible in the blue wing of the observed profiles (see 
Fig.~\ref{fig:specdynI06}, left panel), that the imaging code did apparently not convert into a 
starspot.  The reason for this is that the code assumes the rotating star to host starspots exhibiting no 
intrinsic variability;  if the observed (unfitted) feature was due to such a starspot, we 
would expect a similar signature to show up in the red wing of the observed profiles around phase 0.5, 
i.e., when the putative spot has reached the receding stellar limb.  Since this is not the case here, we 
speculate that the residual unfitted spectral feature is either caused by a small flare or by an 
intrinsically variable starspot.  

Both datasets yield spot occupancies of about 2\% of 
the overall photosphere at both epochs.  This rather low spottedness directly reflects the fact that 
distortions of unpolarised line profiles are rather small.  Note that only spots (or spot groups) with sizes 
comparable to or larger than the resolution element can be reconstructed with DI, tiny isolated spots much 
smaller than our spatial resolution threshold remaining out of reach.  The resulting brightness maps are shown in 
Fig.~\ref{fig:di_maps}. While both maps are roughly similar, they apparently differ by more than a simple phase 
shift (that could result from a small error in the rotation period).  Note that no obvious polar spot is 
reconstructed on V374~Peg.  This result is quite reliable; since V374~Peg is a fast rotator, any strong polar 
structure would produce a clear distortion in the core of the Stokes $I$ profiles, giving them an obvious 
flat-bottom shape.  This is not the case here, as one can see on Fig.~\ref{fig:specdynI05}

Small changes in \vsini\ (within 0.5~\kms), in the inclination angle (within 10\degr), in the local 
profile model (see above) or in the limb-darkening model (linear vs quadratic) produce negligible 
modifications in the reconstructed spot distributions.

\section{Modelling circularly polarised profiles}

\subsection{Model description}
\label{sec:modV}
We use Zeeman-Doppler Imaging (ZDI) to turn our series of rotationally modulated circularly-polarised LSD 
profiles into magnetic field maps at the surface of V374~Peg.  
ZDI is based on the same basic principles as DI except that it aims at mapping a vector field.  
The overall characteristics of the model we use are given below.  

The vector magnetic field at the surface of the star is described using
spherical harmonics expansions for each 
of the field component in spherical coordinates \citep[e.g.,][]{Donati01,
Donati06b}.  Note that this is different 
from the initial ZDI attempts \citep[e.g.,][]{Brown91, Donati97} in which
the field was described as a set of 
independent pixels, containing the components of the field in each surface
grid cell.  The main advantage of this 
new method is that it straightforwardly allows the recovery of a
physically meaningfull magnetic field, that we can 
easily split into its poloidal and toroidal components.  This is of
obvious interest for all studies on stellar 
dynamos.  Another important point is that this new method is very
successfull at recovering simple magnetic field 
structures such as dipoles (even from Stokes $V$ data sets only), while
the old one reportedly failed at such tasks \citep{Brown91, Donati01}.  

As for DI (see Sec.~\ref{sec:modI}), the stellar surface is divided into a
grid of 3~000 pixels on which the magnetic field components are computed
from the coefficients of the spherical harmonics expansion. The
contribution of each grid point to the synthetic Stokes $V$ spectra are
derived using the weak-field approximation: 
\[\]
\begin{equation} 
 V(\lambda) = -g \frac{\lambda_{\rm 0}^2 e}{4 \pi m^2} B_{\rm \ell} \frac{dI}{d\lambda} 
\label{eq:weak}
\end{equation} 
where $I$ is the local unpolarised profile (modelled with the same
gaussian profile as in Sec.~\ref{sec:modI}), $g$ and $\lambda_{\rm 0}$ the
Land\'e factor and the centre rest wavelength of the average LSD line (set
to 1.2 and 700~nm respectively), $e$ and $m$ the electron charge and mass
respectively, and $B_{\rm \ell}$ the line-of-sight projection of the local
magnetic field vector.  Despite being an approximation, this expression
is valid
for magnetic splittings significantly smaller than the local profile width
(prior to instrumental broadening);  given that 2~kG fields yield Zeeman
splittings of only about 2.5~\kms\ in the case of our average LSD profile
(i.e., about twice smaller than the FWHM of the local profile prior to
instrumental broadening), this expression turns out to be accurate enough
for our needs.  This is further confirmed by using model local profiles
computed from Unno-Rachkovsky's equations, with which almost identical
results are obtained.

As for DI, the inversion problem is partly ill-posed.  We make the
solution unique by introducing an 
entropy function describing the amount of information in the reconstructed image, and by looking for the image 
with maximum entropy (minimum information) among all those fitting the data at a given \chisqr\ level.  
The entropy is now computed from the sets of complex spherical harmonics coefficients, rather than from the 
individual image pixels as before. We chose one of the simplest possible form for the entropy:  
\begin{equation}
 S = -\sum_{\ell, m} \ell \left(\alpha_{\ell, m}^2 + \beta_{\ell, m}^2 + \gamma_{\ell, m}^2\right)
\label{eq:ent}
\end{equation}
where $\alpha_{\ell, m}$, $\beta_{\ell, m}$, $\gamma_{\ell, m}$ are the spherical harmonics coefficient of order $(\ell, m)$ describing respectively the radial, non-radial poloidal and toroidal field components. More detail about the process can be found in \citet{Donati06b}.  
In particular, this function allows for negative values of the magnetic field (as opposed to the 
conventional expression of the Shannon entropy).  The precise form of entropy (e.g., the multiplying factor 
$\ell$) has only minor impact on the resulting image.

As mentioned in in Sec.~\ref{sec:modI}, the number of resolved elements on the equator is about 25, indicating 
that limiting the spherical harmonics expansion describing the magnetic field to orders lower than 25 should be 
enough and generate no loss of information.  In practice, no significant change of the reconstructed image is 
observed when limiting the expansion at orders $\leq10$.  All results presented below are obtained assuming 
$\ell\leq10$.

\subsection{Results}
\label{sec:resV}

We present here the results concerning the Aug05 dataset, though already reported in D06, in order to clarify the comparison with the Aug06 observations. The Aug05 data set can be fitted with a purely poloidal field down to $\chisqr=1.1$ (see Fig.~\ref{fig:specdyn05} for the observed and modelled dynamic spectra at this epoch).  No improvement to the fit is obtained when assuming that the field also features a toroidal component.  We therefore conclude that the field of V374~Peg is mostly potential in Aug05 and that the toroidal component (if any) includes less than 4\% of the overall large-scale magnetic field energy.  The field we recover is shown in Fig.~\ref{fig:zdi_maps} (left column) while the corresponding spherical harmonics power spectra are shown in Fig.~\ref{fig:zdi_coeffs} (left column). The reconstructed large-scale field is found to be mostly axisymmetric, with the visible hemisphere being 
mostly covered with positive radial field (i.e., with radial field lines emerging from the surface of the star).  
As obvious from Fig.~\ref{fig:zdi_coeffs}, the dominant mode we recover corresponds to $\ell=1$ and $m=0$ (i.e., 
a dipole field aligned with the rotation axis) and has an amplitude of about 2~kG;  all other modes have 
amplitudes lower than 1~kG.  The maximum field strength at the surface of the star reaches 1.3~kG, while the 
average field strength is about 0.8~kG.   

Similar results are obtained from the Aug06 data, for which the observed spectra are fitted down to $\chisqr=1.0$ 
(see Fig.~\ref{fig:specdyn06}).  Again, we find that the field is mostly axisymmetric, with 80\% of the 
reconstructed magnetic energy concentrating in $m=0$ modes.  The radial field component clearly dominates (75\% 
of the energy content) while the toroidal component is once more very marginal (4\% of the energy content).  
The main excited mode again corresponds to $\ell=1$ and $m=0$, features an amplitude of 1.9~kG and encloses as 
much as 60\% of the overall magnetic energy content.  The maximum and average field strengths recovered at the 
surface of the star are 1.2~kG and 0.7~kG respectively.  

As a result of the limited spatial resolution of the imaging process, small-scale magnetic regions at the surface 
of V374~Peg are out of reach of ZDI.  We can therefore not evaluate the amount of magnetic energy stored at small 
scales and compare it to that we recover at medium and large scales.  We can however claim that scales corresponding 
to spherical harmonics orders ranging from about 5 to 25, to which our imaging process is sensitive, 
contain very little magnetic energy with respect to the largest scales ($\ell\leq5$).   

\begin{figure*}
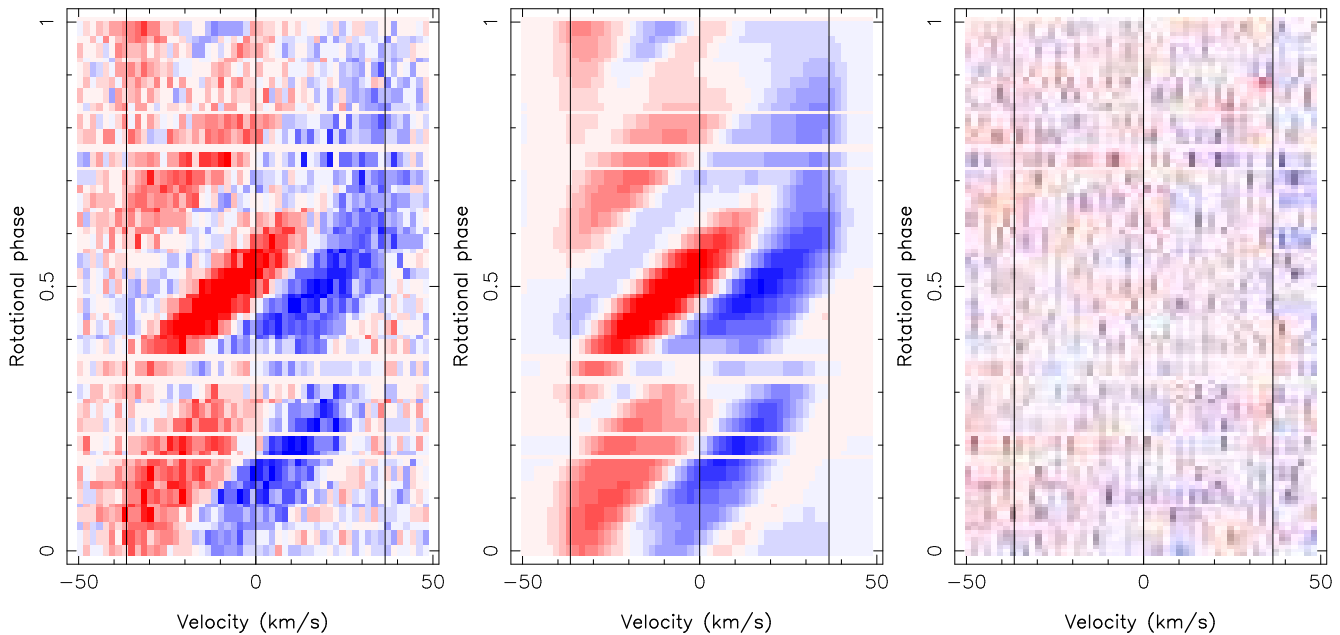

 \center{
 \includegraphics[width=5.8cm]{fig/dynv_aug05b_d.ps}
 \includegraphics[width=5.8cm]{fig/dynv_aug05b_r.ps}
 \includegraphics[width=5.8cm]{fig/dynv_aug05b_res.ps}}
 \caption[]{Observed (left), modelled (middle) and residual (right) Stokes $V$ dynamic spectra of V374~Peg for the 
Aug05 data set (60 spectra). The fit corresponds to $\chisqr=1.1$. The colour scale ranges from --0.4\% to 0.4\% of the continuum level, red and blue respectively standing for positive and negative circular polarization signal.}
 \label{fig:specdyn05}
\end{figure*}

\begin{figure*}
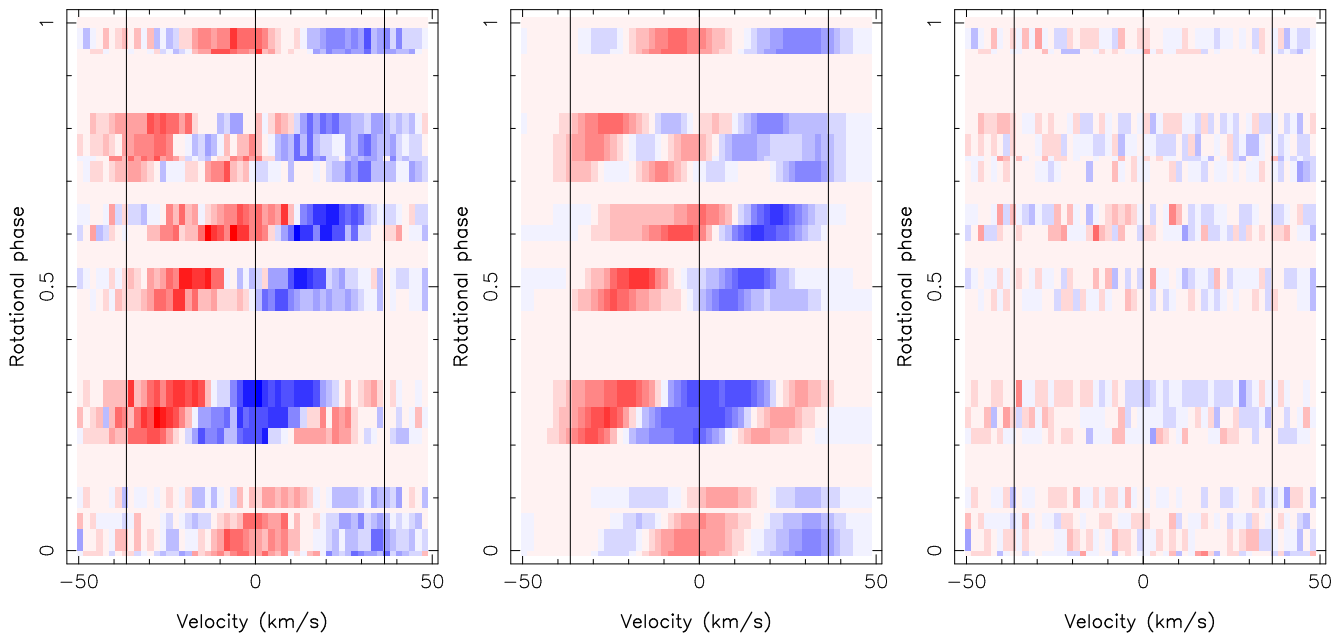

 \center{
 \includegraphics[width=5.8cm]{fig/dynv_aug06_d.ps}
 \includegraphics[width=5.8cm]{fig/dynv_aug06_r.ps}
 \includegraphics[width=5.8cm]{fig/dynv_aug06_res.ps}}
 \caption[]{Same as Fig.~\ref{fig:specdyn05} for the Aug06 data set (21 spectra).  The fit corresponds to 
$\chisqr=1.0$.}
 \label{fig:specdyn06}
\end{figure*}

We can also note that that the magnetic maps recovered from the Aug05 and Aug06 data sets are fairly similar apart 
from a 36\degr\ (0.1 rotation cycle) global shift between both epochs, certainly resulting from a small error in the rotation period. Slight intrinsic differences also seem to be present, e.g., in the exact shape of the positive radial field distribution over the visible hemisphere. A straightforward visual comparison of both images is however not sufficient to distinguish whether these apparent discrepancies simply result from differences in phase coverage at both epochs, or truly reveal intrinsic variability of the magnetic topology between our two observing runs.  To test this, one way is to merge all available data (including those collected in Sep05) in a single data set and to attempt fitting them with a single magnetic structure (and a revised rotation period).  This is done in Sec.~\ref{sec:rot}.  In the meantime, we can nevertheless conclude that the magnetic topology of V374~Peg remained globally stable over a timescale of 1~yr.

\begin{figure*}
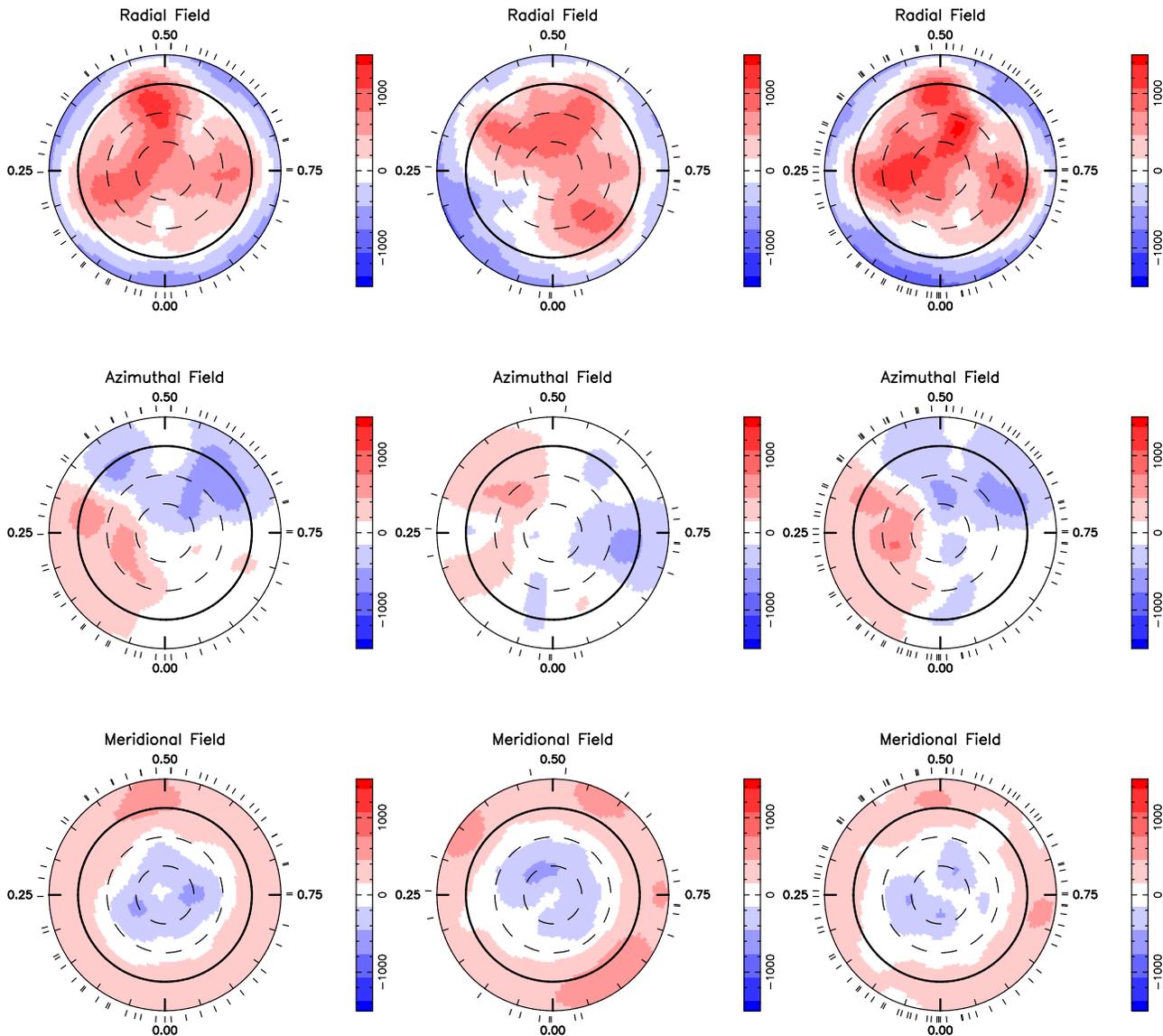

 \center{
 \includegraphics[scale=0.6]{fig/v374peg_V_aug05b.ps}
 \includegraphics[scale=0.6]{fig/v374peg_V_aug06.ps}}
 \includegraphics[scale=0.6]{fig/v374peg_V_0506.ps}
 \caption[]{Surface magnetic field of V374~Peg as derived from our Aug05
(left) and Aug06 (middle) datasets. The last image (right) was recovered
from the complete dataset (Aug05, Sep05, and Aug06 altogether), it
represents the magnetic configuration at an intermediate epoch (cycle 194)
and takes into account the adjustment of the period and
the inclusion of differential rotation (see Sec. 5). The three components
of the field in spherical coordinates are displayed from top to bottom
(flux values labelled in G).  The star is shown in flattened polar
projection down to latitudes of $-30\degr$, with the equator depicted as a
bold circle and parallels as dashed circles.  Radial ticks around each
plot indicate phases of 
observations. }
\label{fig:zdi_maps}
\end{figure*}

\begin{figure}
 \center{
 \includegraphics[scale=0.36]{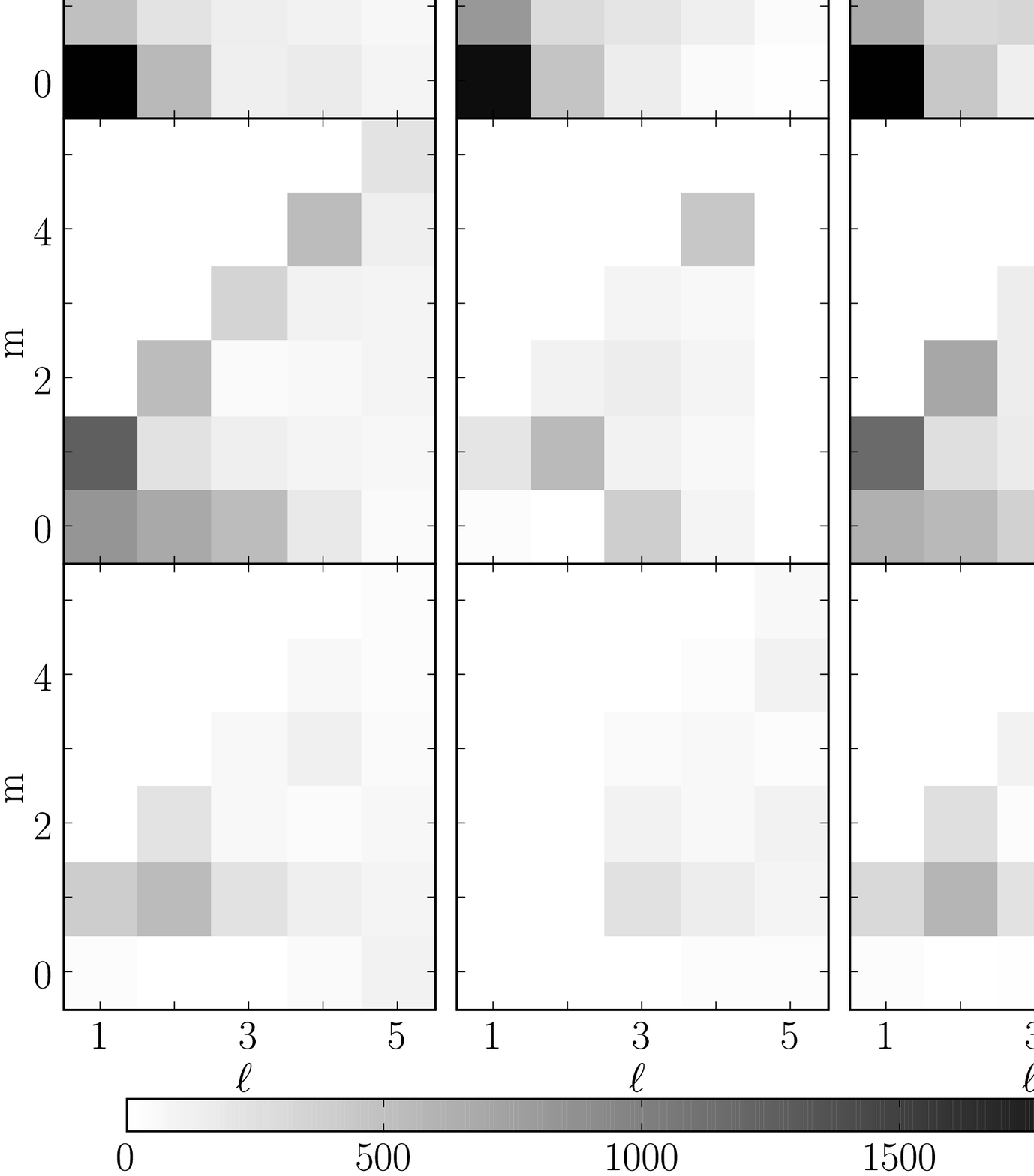}}
 \caption[]{Spherical-harmonics power spectra of the magnetic field maps derived from our Aug05 (left), 
Aug06 (middle) and complete (right) data sets (magnetic fluxes labelled in
G) after adjustment of the period and inclusion of differential rotation
(see Sec.~\ref{sec:rot}).  From top to bottom, the magnetic 
power spectra respectively correspond to the radial poloidal, the
non-radial poloidal and the toroidal field 
components.  Only modes up to order 5 are plotted here (larger order modes containing very little power).  }  
\label{fig:zdi_coeffs}
\end{figure}

\section{Differential Rotation and Magnetic field time-stability}
\label{sec:rot}

To estimate the degree at which the magnetic topology remained stable
over 1~yr, we merge all data together (Aug05, Sep05 and Aug06 datasets)
and try to fit them simultaneously with a single field structure. 
We also assume that the surface of the star is rotating differentially
(with a given differential rotation law, see below) over the
complete period of our observations, and thus that the magnetic topology
is slightly distorted as a result of this surface shear (with respect to
the magnetic topology at median epoch, e.~g. at cycle 194). If the fit to
the complete data set is as good as that to both individual Aug05 and
Aug06 data sets ($\chisqr\simeq1$, see Sec.~\ref{sec:resV}), it would
demonstrate that no variability occurred (except that induced by
differential rotation) over the full observing period.  If, on the other
hand, the fit quality to the complete data set is worse, it would indicate
that intrinsic variability truly occurred, with the fit degradation
directly informing on the actual level of variability.  

% Our ZDI code aims at reconstructing a set of spherical harmonics
% coefficients describing the magnetic field at a median epoch. Thus, for
% each phase of observation the surface of the star is divided into a
% grid,
% and the local field value is computed (from the spherical harmonics
% coefficients) for each pixel. The grid is then sheared by rotating rings
% of constant latitude in order to fit a given differential rotation law.
% Finally synthetic spectra are computed from these grids and compared
% iteratively to the observations untill the maximum entropy solution is
% found.

% In our ZDI code, differential rotation is directly included in the
% response matrix which links the reconstructed quantities (here the
% spherical harmonics decomposition coefficients) to the Stokes profiles
% at
% the various phases of observations (See \citet{Skilling84} for more
% details).  In order to do so, the surface of the star is divided into a
% grid distorted by differential rotation, and the derivatives of the
% spectra with respect to the local field parameters are computed at each
% pixel. The response matrix elements are then computed from these
% derivatives and the appropriate Lengendre polynomials.

To model differential rotation, we proceed as in \citet{Petit02} and
\citet{Donati03}, assuming that the 
rotation rate at the surface of the star is solar-like, i.e., that $\Omega$ varies with latitude $\theta$ as:  
\begin{equation}
\Omega(\theta) = \Omeq - \dOm \sin^2 \theta
\label{eq:drot}
\end{equation}
where $\Omega_{\rm eq}$ is the rotation rate at the equator and $d\Omega$
the difference in rotation rate between the equator and the pole. We use
this law to work out the phase shift of each surface grid cell, at
any given epoch, with respect to its position at median epoch. These phase
shifts are taken into account to compute the synthetic spectra
corresponding to the current magnetic field distribution at all observing
epochs using the model described in Sec.~\ref{sec:modV}.

For each pair of \Omeq\ and \dOm\ values within a range of acceptable
values (given by D06), we then derive, from the complete data set, the
corresponding magnetic topology (at a given information content) and the
associated \chisqr\ level at which modelled spectra fit observations.  By
fitting a paraboloid to the \chisqr\ surface derived in this process
\citep{Donati03}, we can easily infer the magnetic topology that yields
the best fit to the data along with the corresponding differential
rotation parameters and error bars.  

The best fit we obtain corresponds to $\chisqr=1.15$, demonstrating that intrinsic variability between Aug05 
and Aug06 remains very limited, as guessed in Sec.~\ref{sec:resV} by
visually comparing the images 
from individual epochs.  The derived \chisqr\ map (see Fig.~\ref{fig:chisqmap}) features a clear paraboloid 
around the \chisq\ minimum, yielding differential rotation parameters at
the surface of V374~Peg equal to 
$\Omeq=14.09880\pm0.00006$~\rpd\ (corresponding to $P_{\rm rot}=0.445654\pm0.000002$~d) 
and $\dOm=0.0063\pm0.0004$~\rpd.  
Due to the large time gap between the two main observing epochs (about 1yr), the \chisqr\ map exhibits aliases 
on both left and right sides of the minimum, corresponding to shifts of $\sim6\times10^{-4}$~d on \Prot.  
The nearest local minima, located at $\Omeq=14.08110$~\rpd\ and $\Omeq=14.11664$~\rpd, are associated with 
\chisqr\ values of 1.27 and 1.30 respectively, i.e., to $\Delta\chisq$ values of 600 and 740 
respectively;  the corresponding rotation rates can thus be safely excluded.  

In addition to refining the rotation period (error bar shrunk by 2 orders of magnitude with respect to D06), 
we also detect a small but definite surface shear, about 10 
times weaker than that of the Sun. Assuming solid body rotation does not allow to reach \chisqr\ levels 
below 1.5;  this unambiguously demonstrates that the surface of V374~Peg is not rotating as a solid body and 
is truly twisted by differential rotation. The reconstructed magnetic map along with the corresponding 
spherical-harmonics power spectra are shown in the right panels of Figs.~\ref{fig:zdi_maps} and 
\ref{fig:zdi_coeffs}.

The same procedure is used on Stokes $I$ data.  However, the \chisqr\ map we obtain does not feature a clear 
paraboloid, but rather a long valley with no spatially well-defined minimum;  this is not unexpected given 
the fact that spots reconstructed at the surface of V374~Peg do not span a wide range of latitudes (as 
requested for a clean differential rotation measurement, see \citealt{Petit02}) but rather concentrate at
low latitudes.   Using the differential rotation parameters derived from Stokes $V$ data yields a fit 
corresponding to a \sn\ level of about 680, i.e., only slightly lower than that achieved at each individual 
epoch (see Sec.~\ref{sec:resI}).  It suggests that only moderate intrinsic variability occurred in the 
brightness distribution as well, despite the apparent (but weakly significant) differences between the 
two individual images.  The corresponding image (reconstructed from the full data set) is shown in 
Fig.~\ref{fig:di_maps} (right panel).  

\begin{figure}
 \center{
 \includegraphics[angle=270, scale=0.34, clip=true]{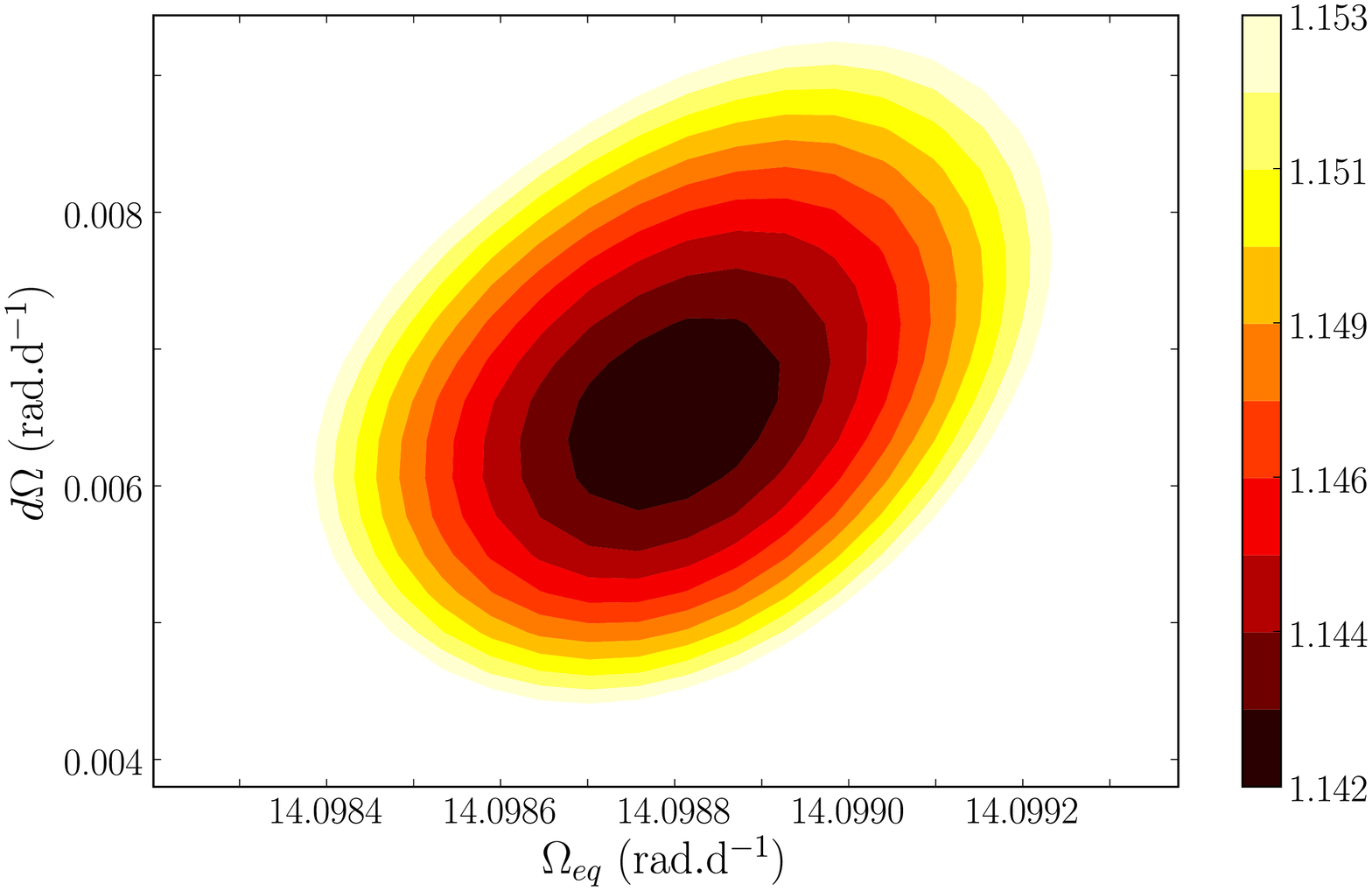}
 \caption[]{Variation of \chisqr\ as a function of the differential rotation parameters \Omeq\ and 
\dOm.  The last colour contour corresponds to a \chisqr\ increase of 1\%, i.e., to an error bar 
of about 7$\sigma$ (for each parameter taken separately). }}
 \label{fig:chisqmap}
\end{figure}

\section{Chromospheric Activity}

We compute the mean and standard deviation profiles of the H$\alpha$, H$\beta$ and \caii\ 
infrared triplet (IRT) lines.  Both H$\alpha$ and H$\beta$ lines are in strong emission whereas the infrared 
triplet is observed mostly in absorption with sporadic emission episodes (due to flares).  We notice that, 
for all studied lines, variability concentrates within $\pm \vsini$ about
the line centre (see Fig.~10), suggesting that activity is mostly located
close to the surface.

For each observation, we compute the emission equivalent width of the chromospheric lines.  The result for 
H$\alpha$ is plotted on Fig.~11 as a function of rotational
phase, H$\beta$ and \caii\ IRT 
exhibiting a very similar behaviour.  We first conclude that all three lines do not undergo pure 
rotational modulation, sometimes displaying very different emission levels at similar phases of successive 
rotation cycles (e.g., phase 0.2, see Fig.~11).  We attribute the basal
emission component 
(in Balmer lines) to quiescent activity and the intrinsically variable emission component to flares.  

Maximum flaring activity is observed around phases 0.20 
and 0.75--1.00, but also, to a lesser extend at phases 0.50--0.60 (see
Fig.~11, bottom panel).  Note that the four photospheric profiles
discarded from our imaging analysis (see Sec.~\ref{sec:obs}) were 
collected on 2005 August 21 at phases 0.72 and 0.78 and on 2005 August 23
at phases 0.19 and 0.23, i.e., at 
the very beginning of the two main flaring episodes we recorded.  

\begin{figure}
 \center{
 \includegraphics[scale=0.45, clip=true]{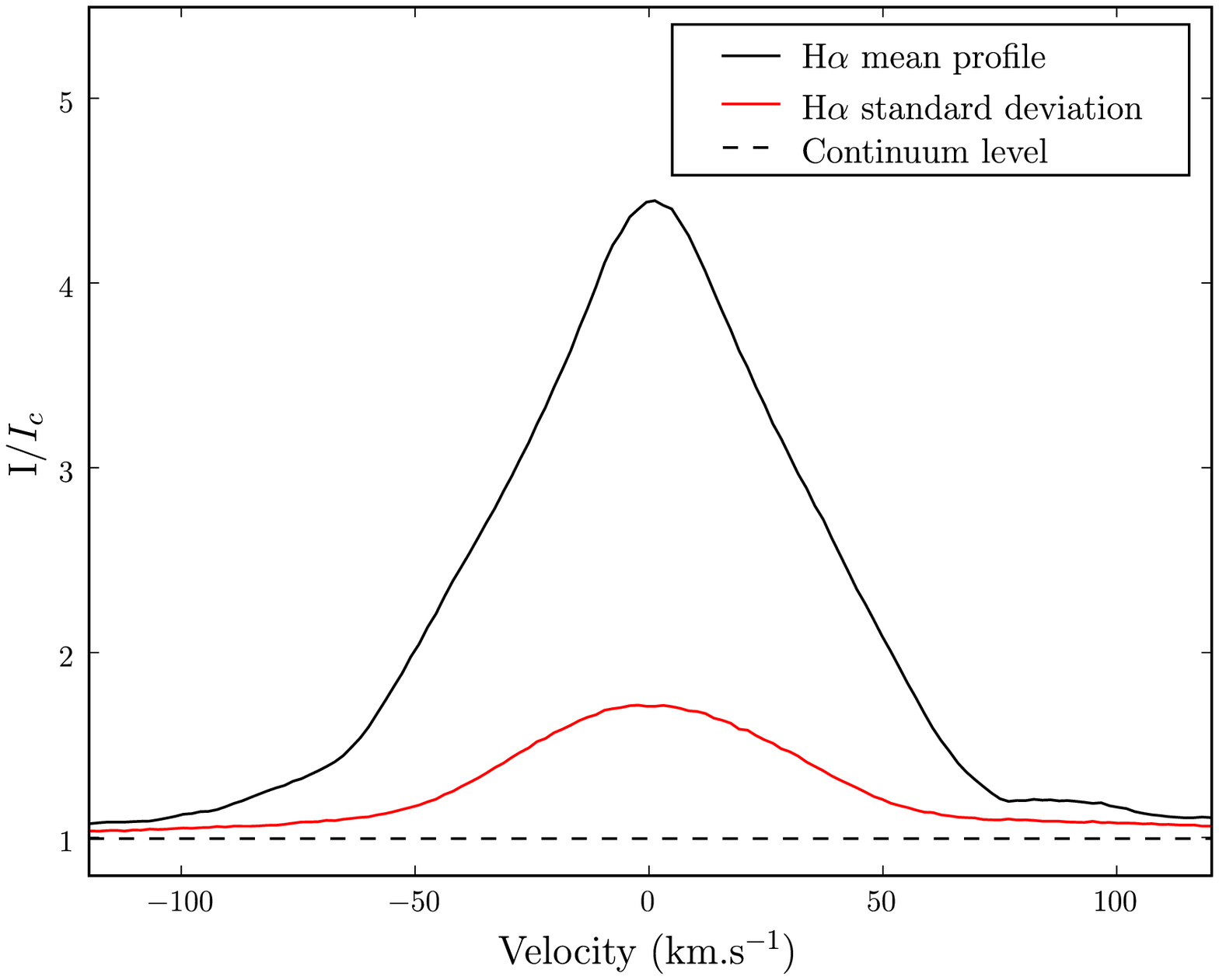}
 \caption[]{H$\alpha$ mean profile (black solid line) and standard deviation (red solid line, shifted by $+1$), 
along with continuum level (dashed line).  All profiles are shown in the stellar velocity rest frame. }}
 \label{fig:Ha1}
\end{figure}

\begin{figure}
 \center{
 \includegraphics[angle=270, scale=0.31, clip=true]{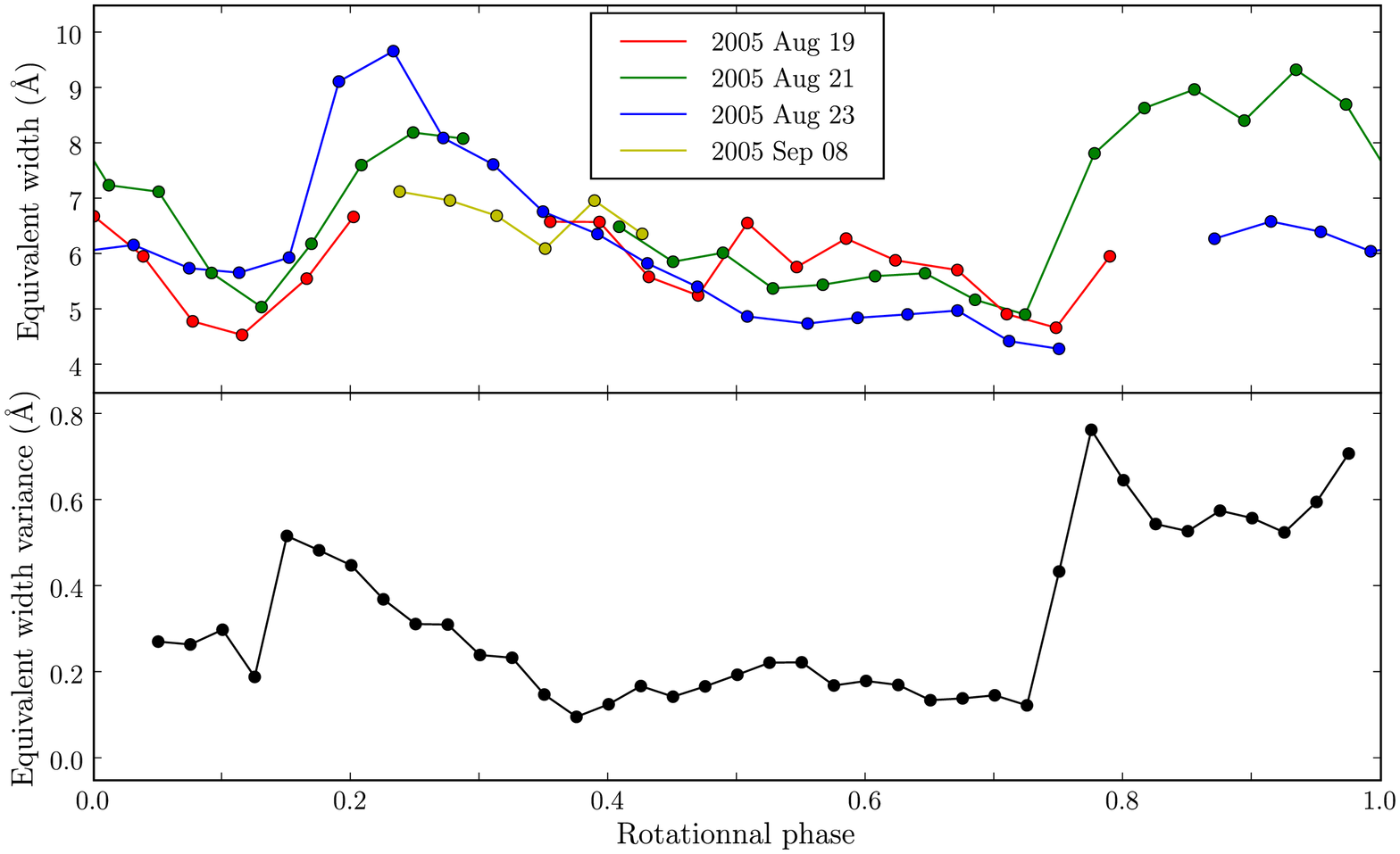}
 \caption[]{Emission equivalent width of H$\alpha$ as a function of the rotational phase in Aug05 and Sep05, consecutive observations are linked together by a solid line (top panel); 
along with associated standard deviation (computed over phase intervals of 0.1~cycle, bottom panel).  }}
 \label{fig:Ha2}
\end{figure}

\section{Discussion and conclusion}
\label{sec:disc}

Spectropolarimetric observations of V374~Peg were carried out with ESPaDOnS and CFHT at 3 epochs over 
a full time span of about 1~yr.  Clear Zeeman signatures are detected in most Stokes $V$ spectra;  weak 
distortions are also observed in the unpolarised profile of photospheric lines.  

Using DI, we find that only a few low-contrast spots are present at the surface of V374~Peg, covering altogether 
2\% of the overall surface.  The spot distribution is apparently variable on a timescale of 1~yr.  Even though 
we may be missing small spots evenly peppering the photosphere \citep[e.g.,][]{Jeffers06}, V374~Peg nevertheless 
appears as fairly different from warmer, partly-convective, active stars, whose surfaces generally host  
large, high-contrast cool spots often covering most of the polar regions at rotation rates as high as that of 
V374~Peg \citep[e.g.,][]{Donati00}. \cite{Barnes04} also report the lack of a cool polar structure for the
M1.5 -- largely but not fully convective -- fast rotator HK~Aqr ($\Prot = 0.431$~d).  A spot occupancy
of about 2\% has been observed for very slow rotators as well \citep{Bonfils07, Demory07}, suggesting 
that spot coverage may not correlate with rotation rate in M dwarfs, as oppposed to what is observed in 
hotter stars \citep[e.~g.,][]{Hall91}.

Using ZDI, we find that the surface magnetic topology of V374~Peg is mostly poloidal and fairly simple, with 
most of the energy concentrating within the lowest-order axisymmetric modes.  This confirms and amplifies the 
previous results of D06.  Again, this is at odds with surface magnetic topologies of warmer active stars, usually 
showing strong and often dominant toroidal field components in the form of azimuthal field rings more or less 
encircling the rotation axes \citep[e.g.,][]{Donati03a}.  As for surface spots, ZDI is mostly insensitive to 
small-scale magnetic regions (and in particular small-scale bipoles) that can also be present at the surface 
of V374~Peg;  we can however claim that very little magnetic energy concentrates in spherical-harmonic terms 
with $\ell$ ranging from 5 to 25.  

We find that the surface magnetic topologies of V374~Peg at epoch Aug05 and Aug06 are very similar apart from 
an overall shift of about 0.1 rotation cycle.  This is confirmed by our success at fitting the complete Stokes 
$V$ data set down to almost noise level with a unique magnetic field topology twisted by a very small amount of 
solar-like differential rotation.  It suggests in particular that magnetic topologies of fully-convective stars 
are globally stable on timescales of at least 1~yr.  Again, this is  highly unusual compared to the case of 
warmer active stars, whose detailed magnetic topologies evolve beyond recognition in timescales of only a few 
weeks \citep[e.g.,][]{Donati03a}.  

The updated rotation period and photospheric shear we find for V374~Peg are respectively equal to 
$0.445654\pm0.000002$~d and $0.0063\pm0.0004$~\rpd respectively;  the corresponding time for the equator to 
lap the pole by one complete rotation cycle is equal to 2.7~yr, about 10 times longer than solar.  We 
speculate that this unusually small differential rotation and long lap time are in direct relationship with 
the long lifetime of the magnetic topology, suggesting that differential rotation is what mostly controls the 
lifetime of magnetic topologies.  New data collected on a smaller time span (a few months) and featuring in 
particular no intrinsic variability (other than that generated by differential rotation) are needed to provide 
an independent and definite confirmation of this result.  

Although the spot distributions at the surface of V374~Peg at epoch Aug05 and Aug06 apparently differ from 
more than a simple phase shift of 0.1 cycle, both data sets are compatible with a unique brightness map 
undergoing moderate intrinsic variability (assuming differential rotation is similar to that derived from 
Stokes $V$ data).  The low spottedness level and weak spot contrast implies that densely sampled data sets at 2 
nearby epochs are needed to confirm this result, derive an independent determination of differential 
rotation parameters from unpolarised spectra and check that spots and magnetic fields trace the same 
atmospheric layers.  

The strong mostly-axisymmetric poloidal field of V374~Peg can hardly be reconciled with the very weak level 
of surface differential rotation in the context of the most recent dynamo models of fully-convective stars.  
While \citet{Dobler06} predicts that rapidly-rotating fully-convective stars should be able to sustain 
strong poloidal axisymmetric fields with the help of significant (though antisolar) differential rotation, 
\citet{Chabrier06} find that these stars should only be able to trigger purely non-axisymmetric magnetic 
topologies if differential rotation is lacking (or remains below a threshold of about 1/3 of the solar 
value, Chabrier, private communication).  Both findings are at odds with our observational results, 
suggesting that our theoretical understanding of dynamo action in fully-convective stars is still 
incomplete.  

We report the detection of strong chromospheric activity in V374~Peg, including quiescent emission 
in Balmer lines along with sporadic flaring events in H$\alpha$, H$\beta$ and the \caii\ IRT.  This 
is noticeably different from what was recently reported by \citet{Berger07} for the M8.5 brown dwarf 
TVLM~513-46546, whose rotationally modulated H$\alpha$ emission identically repeats over several successive 
rotation cycles.  
From the width and variability of emission profiles, we conclude that activity occurs mostly close to 
the stellar surface on V374~Peg, with flares apparently concentrating at specific longitudes.  
However, there is no clear spatial correlation between the photospheric magnetic field, the starspots and 
these chromospheric regions, e.g., strongest magnetic fields being observed when chromospheric activity 
is low.  Simultaneous multiwavelength observations of V374~Peg, including in particular optical 
spectropolarimetry, X-ray, and radio monitoring, could allow us to investigate the 3D magnetosphere of 
fully convective stars (e.g., by extrapolating the photospheric field topology up to the corona), 
and to unravel how photospheric activity relates to chromospheric and coronal activity.  

Finally, we note that our \vsini\ and rotation period estimates imply that $\rstar\sin i=0.32\pm0.01$~\rsun, 
and thus that $\rstar\simeq0.34$~\rsun\ (assuming $i\simeq70\degr$). 
Given the $M_{\rm K}$ magnitude of V374~Peg (equal to $7.0\pm0.3$), 
we can safely derive that V374~Peg has a mass of $0.28\pm0.05$~\msun\ \citep{Delfosse00}.  It implies that 
the radius of V374~Peg is slightly larger than what theoretical models predict for low-mass stars of similar 
masses.  \cite{Ribas06} also reports stellar radii larger than the theoretical predictions for very-low mass 
eclipsing binaries (whose rotation period is very short), whereas the radii of slowly rotating M dwarfs 
match the predicted values \citep[e.~g.,][]{Segransan03}.  According to the phenomenological model of 
\cite{Chabrier07} the presence of a strong magnetic field may inhibit stellar convection and therefore 
lead to larger radii with respect to an inactive star of equal mass.

% Our result thus supports previous similar claims derived from accurate photometric observations of 
% several eclipsing low-mass binaries \citep[e.g.,][]{Ribas06}.  

Future observations are thus needed to explore in more details the magnetic topologies, brightness distributions 
and differential rotation of V374~Peg and other fully-convective stars.  Finding out how these quantities vary 
with mass and rotation rate could in particular show us the way to more successfull simulations of dynamo 
processes in low-mass fully-convective stars.

\section*{ACKNOWLEDGMENTS} 
We thank the CFHT staff for his valuable help throughout our observing runs.  

%% Bibliography

%\bibliography{v374peg}

%\bibliographystyle{mn2e}

\end{document}